\documentclass[preprint,aip]{revtex4}

\usepackage{graphicx}
\usepackage{dcolumn}
\usepackage{bm}

\begin{document}
\title{Influence of the surface termination to the point imaging by a photonic crystal slab with negative refraction}
\author{Sanshui Xiao}
\author{Min Qiu}
\email{min@imit.kth.se}
\author{Zhichao Ruan}
\author{Sailing He}
\affiliation{Joint Research Center of Photonics of the Royal
Institute of Technology (Sweden) and Zhejiang University, Zhejiang
University, Yu-Quan, 310027, PR China, and Laboratory of Optics,
Photonics and Quantum Electronics, Department of Microelectronics
and Information Technology, Royal Institute of Technology (KTH),
Electrum 229, 16440 Kista, Sweden}

\date{\today}

\begin{abstract}
Point imaging by a photonic crystal slab due to the negative
refraction is studied theoretically. By investigating the transfer
function of the imaging system, the influence of the surface
termination to the imaging quality is analyzed. It is shown that
an appropriate surface termination is important for obtaining an
image of good quality.

\end{abstract}

 \maketitle



A photonic crystal  \cite{Yab1987P1,John1987P1,Joannopoulos} is an
artificial structure which has a periodic arrangement of
dielectric or metallic materials. Photonic crystals have been
extensively studied for their unique optical properties. For
instance, they may provide a possibility of forbidding light
propagation within a frequency band, i.e., a photonic bandgap. So
far, most of the studies in the literature are focused on this
optical bandgap property and its applications.

As another equally important property, unconventional ultra-strong
dispersions, e.g., the superprism effect, exist in some photonic
crystals in some frequency regions. Such an ultra-strong
dispersion was firstly reported by Lin \emph{et al.}, and
demonstrated experimentally in the millimeter-wave spectrum
\cite{Lin1996P1}. Kosaka \emph{et al.} then demonstrated the
superprism effect in a highly dispersive photonic microstructure
(a complex  ``autocloned'' photonic crystal) at optical
wavelengths \cite{Kosaka1998P1}. These unusual properties provide
an exciting possibility for obtaining microphotonic and
nanophotonic devices that can focus, disperse, switch, and steer
light. For example, the superprism effect can be used in light
collimating \cite{Kosaka1999P1} and wavelength multiplexing
\cite{Wu2002P1}. Recently, it is found that in some frequency
regions some photonic crystals can also refract light as if they
have a negative refractive index
\cite{Qiu2003P1,Notomi2000P1,Cubukcu2003P1}. This has many
potential applications such  as self-focusing and imaging
\cite{Luo20021P1,Parimi2003P1,Li2003P1}. Different from
conventional lenses, in some photonic crystals evanescent waves
could be amplified due to the existence of surface modes
\cite{Luo2003P1}, as well as in a left-handed material
\cite{Pendry2002P1}. This is vital for subwavelength imaging.

In the present letter, we study point imaging by a two-dimensional
photonic crystal slab due to the negative refraction, in
particular,  the influence of the surface termination of the
photonic crystal slab to image quality.

The two-dimensional photonic crystal we considered here is a
triangular lattice of air holes in a dielectric material
$\epsilon=12.96$, with a lattice constant $a$ and a hole radius
$r=0.4 a$. Only the transverse magnetic (TM) modes are considered
here. For waves at a low frequency (or long wavelength), the
material modulation of the periodic structure does not influence
much the wave propagation. For waves at higher frequencies,
propagation in the photonic crystal is complicated. However,
Bloch-Floquet waves (eigen modes in the photonic crystal) travel
through the photonic crystal with a definite propagation direction
despite the presence of scattering \cite{Sakoda}. To visualize and
analyze light propagation in a  photonic crystal, an equal
frequency surface (EFS) \cite{Notomi2000P1} in $\textbf{k} $ space
of the photonic bands can be introduced, where gradient vectors
give the group velocities of the photonic modes. For the photonic
crystal considered here, the shape of the EFS corresponding to a
frequency in the second band is almost circular
\cite{Notomi2000P1}. One can then define an effective refractive
index from the radius of the EFS with Snell's law
\cite{Qiu2003P1,Notomi2000P1,Cubukcu2003P1} and use it to describe
the light refraction in  the photonic crystal. To assure all-angle
negative refraction, here we choose the frequency
$\omega=0.30(a/\lambda)$, at which the effective refractive index
of the  photonic crystal is $-1$. It should be noted that the
behavior in a photonic crystal with $n_{eff}=-1$ is quite
different from that in a left-handed
material(LHM)\cite{Shelby2001,Smith2000P1,Pendry2000P1}. For a LHM
with $n=-1$, light can go through an air-LHM interface without
reflection \cite{Pendry2000P1}. One may observe a quite different
behavior at an interface between air and a photonic crystal even
when the effective refractive index of the  photonic crystal is
$n_{eff}=-1$. For example, the transmission from air into such a
photonic crystal is nearly zero even at a normal incidence when
the normal of the photonic crystal surface is along $\Gamma K$
direction (the reason is that the incident plane wave has an even
symmetry while the Bloch-Floquet wave in the photonic crystal has
an odd symmetry).

\begin{figure}[h]
\includegraphics[width=3.0in]{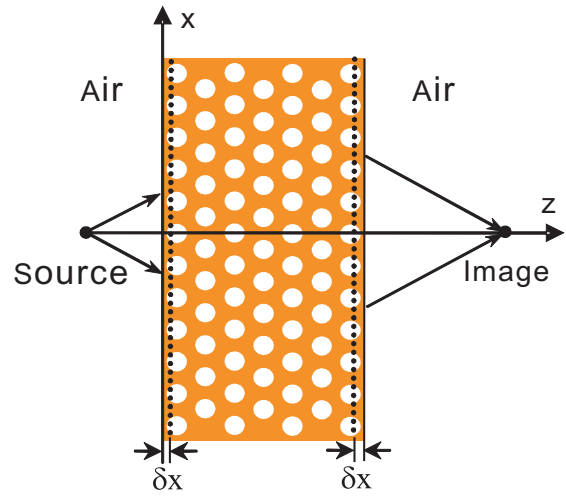}
\caption{\label{slab}Schematic diagram for an imaging system
formed by a photonic crystal slab. The position of the surface
termination is denoted by $\delta x$.}
\end{figure}

\begin{figure}[h]
\includegraphics[width=3.0in]{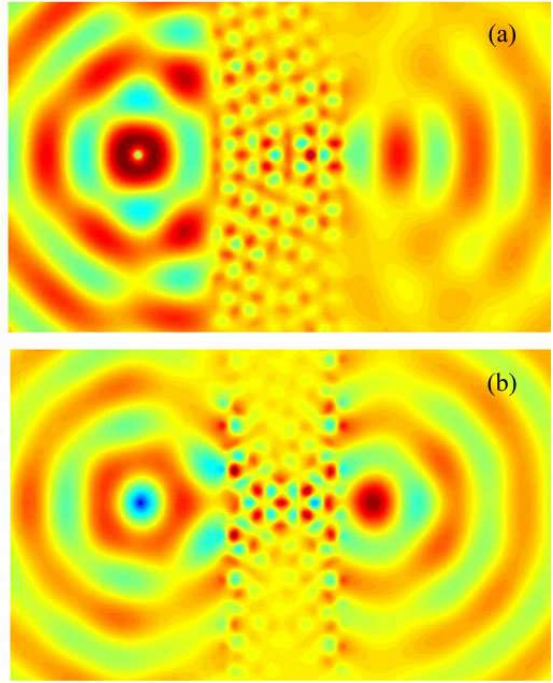}
\caption{\label{image} The snapshots of the electric field of a
point source and its image formed by a photonic crystal slab with
surface termination of (a)$\delta x=0$; (b)$\delta x=0.2 a$.}
\end{figure}

Consider an imaging system composed by a symmetric photonic
crystal slab (with a thickness of seven rows of air holes), which
is surrounded by air. The imaging system with coordinates is shown
in Fig. \ref{slab}, where $\delta x$ is the surface termination of
the  photonic crystal slab at each interface. When $\delta x=0$,
the distance between the left boundary of the photonic crystal
slab and the left boundary of the first column of circles is
$0.1a$. The normal of the slab interface is along the $\Gamma M$
direction. Numerical simulations are performed using the
finite-difference time-domain (FDTD) method \cite{Yee1966P1} with
a boundary treatment of perfectly matched layers
 \cite{PML2D}. A point source of continuous wave is
placed at the left side of the  photonic crystal slab. Figure
\ref{image}(a-b) give the snapshots of the electric field for
$\delta x=0$ and $\delta x=0.2a$, respectively. Focused images are
observed on the right side of the  photonic crystal slab for both
positions of surface termination. The simulations clearly
demonstrate the negative refraction in such a photonic crystal
slab. However, in Fig. \ref{image}(a), the image is relatively
blurred, which indicates that the reflection at the slab interface
is probably quite high. The case of $\delta x=0.2a$ (corresponding
to Fig. \ref{image}(b)) has a better image than the case of
$\delta x=0$ (Fig. \ref{image}(a)). The position of the surface
termination influences quite much the reflection at the interface
between air and the photonic crystal (and consequently the image
quality). Note that it happens that the source and the image in
Fig. \ref{image} have a $\pi$-difference in phase for both
positions of surface termination. In general, an arbitrary phase
shift may be achieved by e.g. choosing an appropriate thickness of
the photonic crystal slab  \cite{Luo20021P1, Li2003P1}.

\begin{figure}[h]
\includegraphics[width=3.0in]{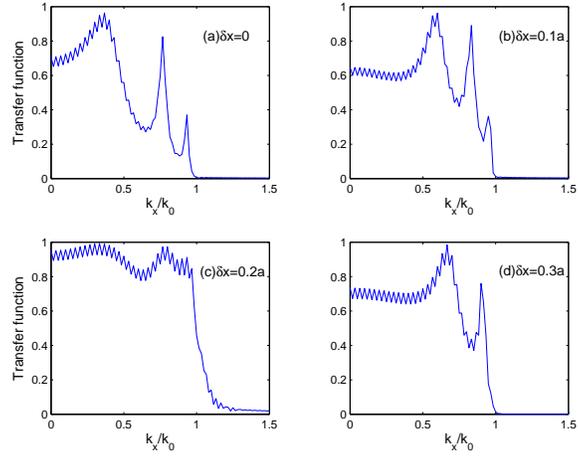}
\caption{\label{trans} Transfer functions of of a photonic crystal
slab (as an imaging system)  for different surface terminations.
$(a)\delta x=0; (b)\delta x=0.1 a; (c) \delta x=0.2 a;(d) \delta
x=0.3 a$.}
\end{figure}

Since the photonic crystal slab is terminated with air at both
sides, only one propagating mode exists at the right side of the
photonic crystal slab for each incident plane wave component.
Thus, we can apply the transfer function method to analyze the
influence of the surface termination to the image quality (see
e.g. \cite{Shen2003P1}). Utilizing the discrete Fourier transform
algorithm and the FDTD method, we can obtain the transfer function
for such a  photonic crystal slab. Figure \ref{trans}(a-d) give
the transfer functions of the imaging system with surface
termination at $\delta x=0$, $\delta x=0.1a$, $\delta x=0.2a$, and
$\delta x=0.3a$, respectively. From Fig. \ref{trans} one sees that
the transmission through the photonic crystal slab is strongly
angular dependent, which is quite different from the case of a LHM
($n=-1$) slab in air. This is mainly due to the angular-dependent
coupling coefficient at the interface between air and the photonic
crystal. Figure \ref{trans} also shows that the transmission
depends on the surface termination of the  photonic crystal slab.
Compared with the results for other surface terminations, the
transmission is relatively large when $\delta x=0.2a$. From Fig.
\ref{trans}(b) one sees that the transfer function for $\delta
x=0.2a$ is relatively flat at all propagating angles, which is
essential for obtaining an image of good quality. From Fig.
\ref{trans} one can also see that there exit some peaks in each
transfer function. The peak positions vary for different surface
terminations. This is probably due to the Fabry-Perot resonant
effect since the effective width of the photonic crystal slab will
decrease as the surface termination of the photonic crystal slab
increases.

\begin{figure}[h]
\includegraphics[width=3.0in]{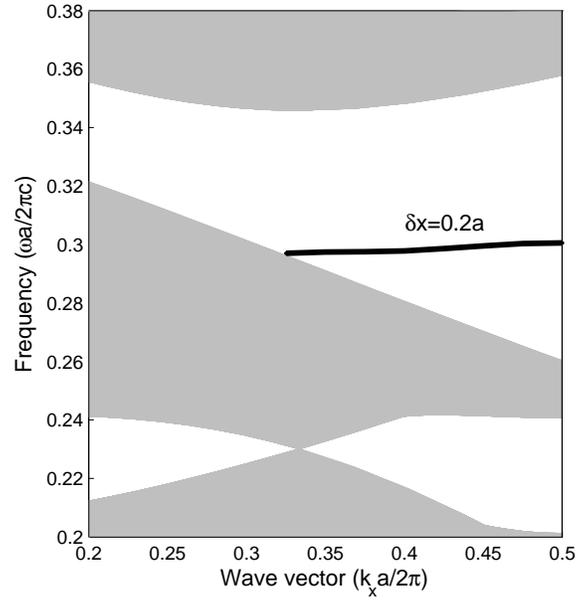}
\caption{\label{surface} The surface mode (solid line) of the
photonic crystal for the termination factor $\delta x=0.2a$. The
shaded regions are the photonic band structures projected along
the $\Gamma K$ direction.}
\end{figure}

Surface modes are generally supported by appropriate crystal
termination surfaces. The surface mode (solid line) of the
photonic crystal along $\Gamma K$ direction with surface
termination $\delta x=0.2a$ is shown in Fig. \ref{surface}. It is
obvious that the surface mode approaches a {\it flat} line near
the working frequency, $\omega=0.30(a/\lambda)$. It has been shown
in \cite{Luo2003P1} that the amplification of evanescent waves
relies on resonant coupling mechanisms to surface photonic bound
states. From Fig. \ref{trans} (c), it is also confirmed that, for
$k_x/k_0>1$ (corresponding to the evanescent components of the
source profile), the transmission of evanescent waves are much
larger than those for other surface termination, with the help of
the surface mode at the working frequency. It is important to
obtain high imaging quality.

In summary, we have studied the point imaging for a slab of
photonic crystal with negative refraction. The influences of the
surface termination to the transmission and the imaging quality
have been analyzed. The transfer function for the imaging system
is obtained by combining the discrete Fourier transform algorithm
with the FDTD method. Our simulation results have shown that the
transmission is strongly angular dependent even when the effective
refractive index of the photonic crystal is matched with the index
of the air. It is also shown that the surface termination plays a
key role in obtaining an image of relatively good quality.

This work was supported by the Swedish Foundation for Strategic
Research (SSF) on Photonics, the Swedish Research Council (VR)
under project 2003-5501, and the National Natural Science
Foundation of China under key project (90101024) and project
(60378037).

\newpage
\section{Figure captions}
\textbf{Figure 1}: (Color Online) Schematic diagram for an imaging
system formed by a photonic crystal slab. The position of the
surface termination is denoted by $\delta x$.

\textbf{Figure 2}: (Color Online) The snapshots of the electric
field of a point source and its image formed by a photonic crystal
slab with surface termination of (a)$\delta x=0$; (b)$\delta x=0.2
a$.

\textbf{Figure 3}: (Color Online) Transfer functions of of a
photonic crystal slab (as an imaging system)  for different
surface terminations. $(a)\delta x=0; (b)\delta x=0.1 a; (c)
\delta x=0.2 a;(d) \delta x=0.3 a$.

\textbf{Figure 4}:  The surface mode (solid line) of the photonic
crystal for the termination factor $\delta x=0.2a$. The shaded
regions are the photonic band structures projected along the
$\Gamma K$ direction.

\newpage
\clearpage
\newpage
\clearpage
\newpage
\clearpage
\newpage
\clearpage

%
%
%
%

\end{document}